# Adaptive Hybrid Optical–RF Transmission Based on Deep Reinforcement Learning for Ultra-Reliable Low-Latency Communication

Stella N. Arinze
*Dept. of Electrical and Electronic*
*Faculty of Engineering*
*Enugu State University of Sci. & Tech.*
Agbani, Nigeria
ndidi.arinze@esut.edu.ng

Isidore U. Uche
*Dept. of Electrical and Electronic*
*Faculty of Engineering*
*Enugu State University of Sci. & Tech.*
Agbani, Nigeria
isidore.uche@esut.edu.ng

Augustine O. Nwajana
*School of Engineering*
*Faculty of Engineering & Science*
*University of Greenwich*
London, UK
a.o.nwajana@greenwich.ac.uk

***Abstract*—Ultra-Reliable Low-Latency Communication (URLLC) demands sub-millisecond latency and reliability above 99.999%, which are difficult to guarantee under dynamic wireless and atmospheric conditions. Although free-space optical (FSO) links provide high-capacity transmission, their performance degrades under turbulence and fog, while radiofrequency (RF) links offer robustness but limited bandwidth. This paper presents a reinforcement learning-based hybrid optical–RF framework that adaptively selects transmission modes to satisfy stringent URLLC requirements. The hybrid decision process is modeled as a Markov Decision Process and optimized using a Deep Q-Network (DQN) controller. Extensive simulations under varying attenuation and fading scenarios demonstrate up to 35% latency reduction and 25–30% throughput improvement compared to RF-only and static hybrid schemes, while maintaining reliability above 99.999%. A laboratory-scale testbed operating at 1550 nm and 28 GHz further validates the approach, showing less than 6% deviation between simulated and measured results. The proposed framework demonstrates the practical feasibility of intelligent hybrid transmission for mission-critical next-generation wireless networks.**

## I. INTRODUCTION

URLLC is a key service category in beyond-5G and emerging 6G wireless networks, supporting mission-critical applications such as industrial automation, autonomous systems, and remote medical procedures. URLLC requires end-to-end latency below 1 ms and reliability exceeding 99.999%, which remain challenging under wireless fading, interference, and spectrum congestion conditions [1]–[2]. FSO communication has emerged as a high-capacity complementary technology to RF transmission due to its large bandwidth and immunity to electromagnetic interference. However, FSO links are highly sensitive to atmospheric turbulence, fog, and misalignment effects that degrade link availability [3]. In contrast, RF links provide robust connectivity under adverse weather conditions but suffer from limited bandwidth and spectrum scarcity [3]–[5]. Hybrid RF/FSO architectures have therefore been proposed to exploit the complementary characteristics of both technologies through dynamic switching or joint transmission strategies [4],[6].Reinforcement learning (RL) has recently gained attention as an intelligent decision-making framework for optimizing resource allocation, spectrum management, and latency-aware transmission control in URLLC systems [7]. RL-based hybrid RF/FSO optimization has also been explored in integrated wireless–optical networks to enhance reliability under dynamic environmental conditions [8]. Nevertheless, most existing hybrid systems rely on static threshold-based switching mechanisms that fail to adapt efficiently to rapidly varying channels, resulting in suboptimal performance [9]. Adaptive transmission strategies based on machine learning offer a promising solution for real-time link selection and reliability enhancement in hybrid networks [10].

Despite these developments, limited work has addressed reinforcement learning-based hybrid optical–RF architectures specifically designed to guarantee stringent URLLC latency and reliability requirements. To address this gap, this paper proposes a reinforcement learning-based hybrid optical–RF framework in which an intelligent agent dynamically selects transmission modes according to instantaneous channel conditions and queue status. The proposed system jointly optimizes latency, reliability, and spectral efficiency, and its performance is validated through comprehensive simulations and experimental prototype implementation.

## II. THEORETICAL CIRCUIT CONFIGURATION

URLLC is one of the primary service categories introduced in 5G and extended toward beyond-5G and 6G systems, as illustrated in Fig. 1. URLLC aims to achieve end-to-end latency below 1 ms and reliability exceeding 99.999%, enabling mission-critical applications such as industrial automation, autonomous transportation, and remote medical procedures [11]–[14]. These stringent requirements are achieved through shortened transmission intervals, mini-slot scheduling, multi-connectivity transmission, adaptive coding, and intelligent resource management. However, meeting ultra-low latency and ultra-high reliability simultaneously remains challenging because latency reduction and reliability enhancement mechanisms often conflict [15].

FSO communication offers high-capacity transmission and immunity to electromagnetic interference, while RF communication provides robust connectivity under adverse weather conditions [17]. Nevertheless, FSO links are highly sensitive to atmospheric turbulence, fog, and misalignment, whereas RF links suffer from spectrum limitations and bandwidth constraints. The Gamma–Gamma model is used for FSO links due to its suitability for moderate-to-strong atmospheric turbulence, while Rayleigh and Rician models are adopted for RF channels to represent non-line-of-sight and line-of-sight propagation conditions, respectively. Hybrid RF/FSO architectures have therefore been proposed to exploit the complementary characteristics of both technologies, improving reliability and throughput compared to single-link systems [18]. Fig. 2 illustrates the hybrid RF/FSO switching architecture, where RF and optical links operate in parallel between source and destination. The switching controller continuously monitors channel conditions and dynamically selects RF-only, FSO-only, or cooperative transmission modes. When optical channel quality is favorable, FSO transmission is prioritized to exploit high bandwidth. Under degraded optical conditions, traffic is redirected to the RF link to maintain reliability. In cooperative mode, simultaneous transmission across both links enhances diversity gain and link availability.

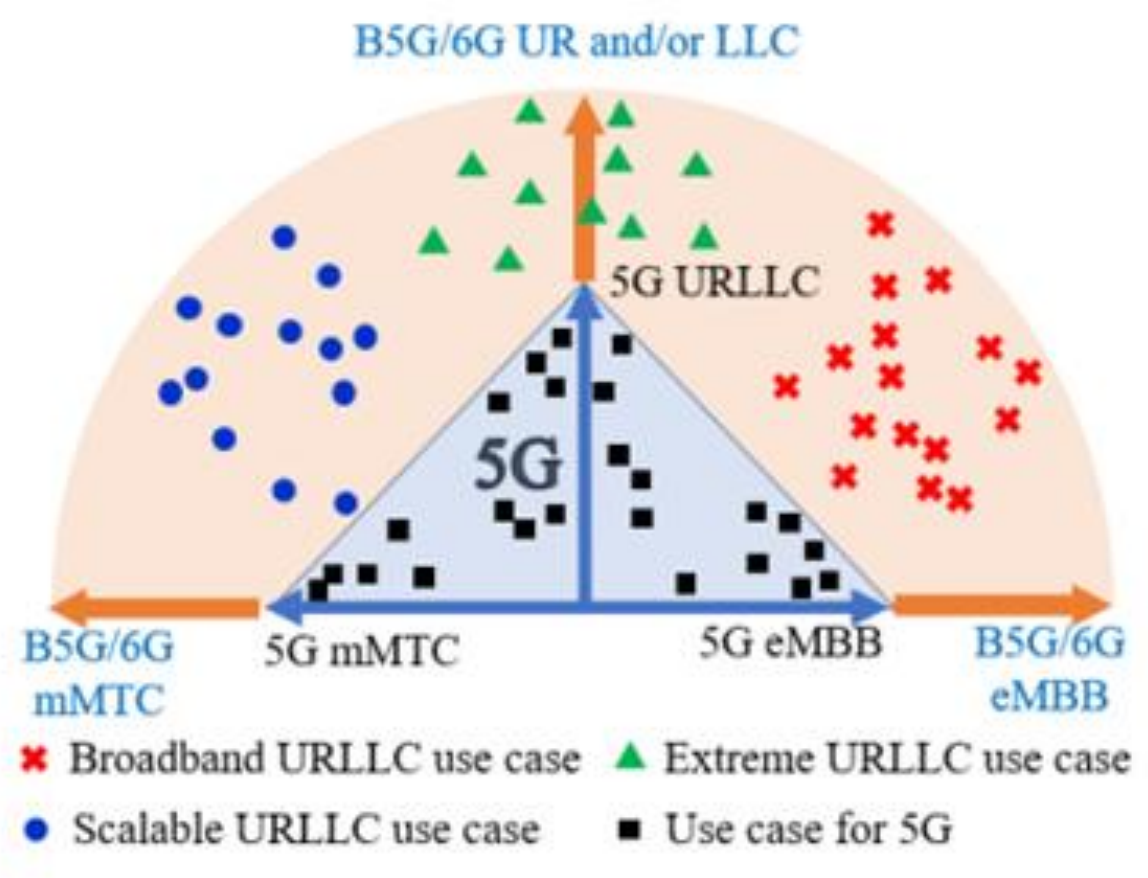

Fig. 1. The 5G service triangle [16].

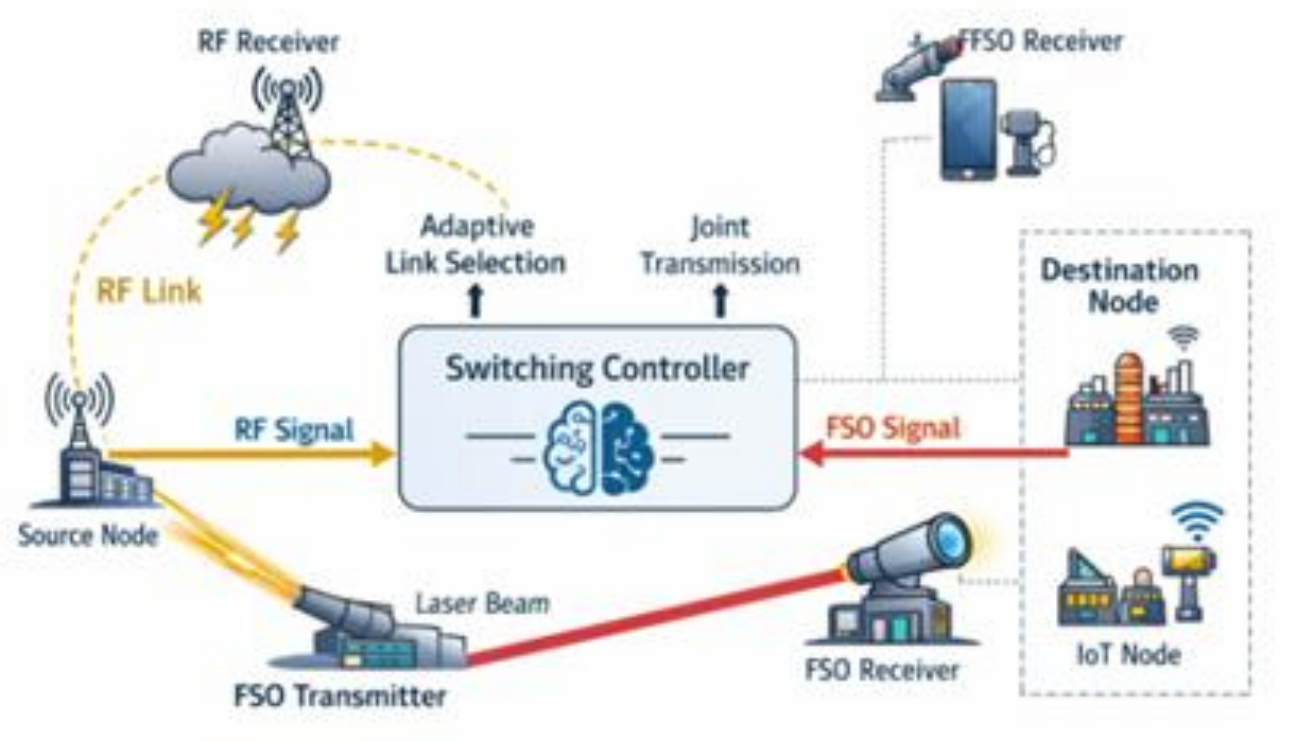

Fig. 2. Hybrid RF/FSO Switching Architecture.

To enable intelligent adaptive switching, the hybrid transmission decision problem is formulated as a Markov decision process (MDP), a standard reinforcement learning framework widely adopted for wireless optimization problems [19]. The MDP is defined as (1) where S, A, P and R denote the state space, action space, state transition probability and reward function, respectively. In the proposed framework, the system state at time slot t is defined as (2) representing instantaneous RF channel quality, optical channel quality, queue length, and packet delay. The action space is defined as (3) corresponding to RF-only, FSO-only, and hybrid cooperative transmission modes. Reinforcement learning-based adaptive hybrid link selection has been shown to improve communication performance and link reliability in hybrid wireless-optical communication systems. To jointly optimize latency and reliability, this work introduces a proposed latency-aware reward formulation in (4) where $D_t$ denotes transmission delay or latency, $P_{Loss,t}$ represents packet loss probability, $R_t$ denotes reliability level, and α. β, γ are weighting coefficients controlling the latency-reliability trade-off (α + β + γ= 1, set as 0.4, 0.3, 0.3 respectively). The optimal transmission policy is obtained by maximizing the expected cumulative discounted reward [20]–[21], a standard RL optimization objective (5). The URLLC requirements are defined as (6). Through this formulation, the reinforcement learning agent learns adaptive hybrid transmission strategies capable of dynamically responding to channel variations and traffic conditions while satisfying URLLC latency and reliability requirement.

$$M = (S, A, P, R) \quad (1)$$

$$s_t = [\gamma RF(t), \gamma FSO(t), q(t), d(t)] \quad (2)$$

$$a_t \in (RF, FSO, HYB) \quad (3)$$

$$r_t = -\alpha \frac{D_t}{D_{max}} - \beta p_{Loss,t} + \gamma R_t \quad (4)$$

$$\pi^* = \arg\max_{\pi} E\left[\sum_{t=0}^{\infty} \partial^t r_t\right] \quad (5)$$

$$D_t \leq 1ms, R_t = 1 - P_{Loss,t}, \geq 1 - 10^{-5} \quad (6)$$

## III. METHODOLOGY

The FSO communication subsystem is modeled using the OptiSystem simulation environment to achieve realistic optical wireless transmission behavior. Fig. 3 illustrates the implementation of the FSO communication subsystem in the OptiSystem simulation environment, representing the optical transmission chain used in the hybrid optical–RF URLLC framework. At the transmitter side, a Pseudo-Random Bit Sequence (PRBS) generator produces the digital data stream representing URLLC traffic. The generated bit sequence is converted into electrical pulses using an NRZ pulse generator, which then drives the optical modulation process. A continuous-wave (CW) laser source operating at a wavelength of 1550 nm provides the optical carrier signal that is modulated by the electrical data using an optical modulator. The modulated optical signal propagates through the free-space optical channel, which incorporates realistic atmospheric effects such as Gamma–Gamma atmospheric turbulence, weather-dependent attenuation, and pointing error impairments. These channel effects model the degradation encountered in practical terrestrial optical wireless communication links. Additional attenuation blocks are included to represent environmental losses caused by fog, rain, and beam divergence during propagation. At the receiver side, the optical signal is detected using an avalanche photodiode (APD) photodetector, which converts the received optical signal into an electrical signal while accounting for photodetector noise characteristics. The detected signal is processed through a low-pass Bessel filter to remove high-frequency noise components and recover the transmitted baseband signal. Performance evaluation blocks then compute important optical communication metrics, including bit-error rate (BER) and optical signal-to-noise ratio (OSNR), which are displayed in the simulation results panel. These performance metrics are exported to the MATLAB system-level simulation environment and subsequently used as input parameters for the reinforcement learning–based hybrid transmission controller. Table I shows the Simulation Parameters of the Optical Subsystem.

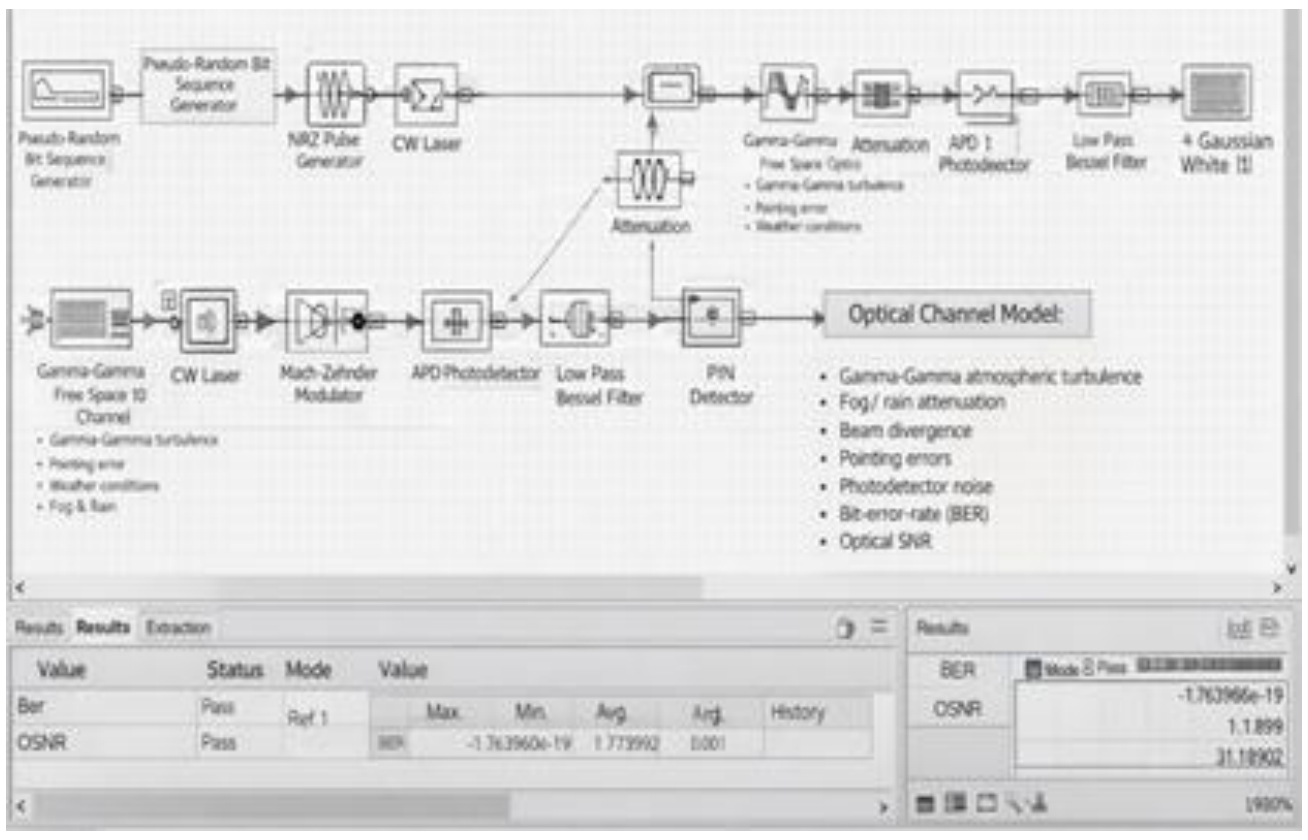

Fig. 3. Optical simulation of free-space communication.

TABLE I. OPTICAL SUBSYSTEM SIMULATION PARAMETERS

| ***Parameter*** | ***Value*** |
|---|---|
| Optical wavelength | 1550 nm |
| Optical transmit power | 10 dBm |
| Propagation distance | 2 km |
| Atmospheric attenuation | 2–20 dB/km |
| Turbulence model | Gamma–Gamma |
| Refractive index structure constant $C_n^2$ | $10^{-14} m^{-2/3}$ |
| Beam divergence angle | 2 mrad |
| Pointing error variance | 1 μrad |
| Receiver type | Avalanche Photodiode |
| Avalanche Photodiode responsivity | 0.9 A/W |
| Electrical filter | Bessel lowpass |
| Optical bandwidth | 1 GHz |

Following the optical subsystem modeling in the OptiSystem environment, the RF communication subsystem and reinforcement learning control architecture are implemented in MATLAB Simulink shown in Fig. 4 to enable intelligent adaptive hybrid transmission. The architecture in figure 4 is organized into five functional layers: traffic generation, RF subsystem, optical interface, reinforcement learning environment, and performance evaluation. The URLLC Traffic Generator produces mission-critical packet data with strict latency constraints. It simulates short-packet arrivals using a stochastic process and maintains a queue buffer representing waiting packets. The output of this block is a digital bit stream that is simultaneously available to both RF and optical transmission paths. The generator also provides queue length and delay information to the reinforcement learning state module. The RF Transmission Subsystem models the wireless communication chain. The incoming bit stream is modulated using a Rectangular QAM Modulator. The signal then passes through a fading channel model representing Rayleigh or Rician propagation conditions, followed by additive white Gaussian noise. At the receiver side, demodulation and error calculation blocks compute RF Signal-to-Noise Ratio and RF Bit Error Rate. These performance metrics are forwarded to the State Metrics block and the reinforcement learning environment. The optical transmission subsystem is represented through the OptiSystem interface block. Although optical physical modeling occurs externally in OptiSystem, its output metrics are imported into Simulink. These include Optical Signal-to-Noise Ratio, atmospheric attenuation, turbulence index, and optical BER. These values are fed into the State Metrics block, ensuring that the reinforcement learning controller has real-time awareness of optical channel conditions. Optical performance metrics (OSNR and BER) are exported from OptiSystem and combined with RF and traffic parameters in MATLAB to form the system state. The overall data flow and synchronization between subsystems are illustrated in Fig. 5. The State Metrics block aggregates all relevant system parameters (OpticalSNR, FogLevel, TurbulenceIndex, QueueLength, RFSNR) into the state vector . This state vector (s(t)) is supplied to the RL Environment block, which contains the Switching Controller implemented using a deep reinforcement learning algorithm such as DQN or PPO. The controller selects a transmission action based on the observed state. Available actions include RF-only transmission, FSO-only transmission, or hybrid RF–FSO transmission. The selected action is passed to the Hybrid Switch Router block, which configures how traffic is distributed between the RF and optical subsystems. The Reward Function block evaluates the outcome of each transmission decision. The reward is computed using system-level metrics such as throughput, outage probability, and delay. The reward signal is fed back into the RL agent, which updates its action-value function and policy parameters. An Experience Replay mechanism ensures stable learning by storing and sampling past state-action transitions. The Performance Evaluation block computes overall system metrics including throughput, latency, and dependability metrics. These outputs are used for convergence analysis and comparison with baseline transmission strategies. Table II shows the Simulation Parameters of the RF Subsystem and Reinforcement Learning. The system performance is evaluated under varying atmospheric attenuation (2–20 dB/km) and RF SNR (0–30 dB) to capture different channel conditions. Increased attenuation degrades FSO performance, while severe fading affects RF links, and the proposed hybrid framework adaptively maintains reliability and low latency.

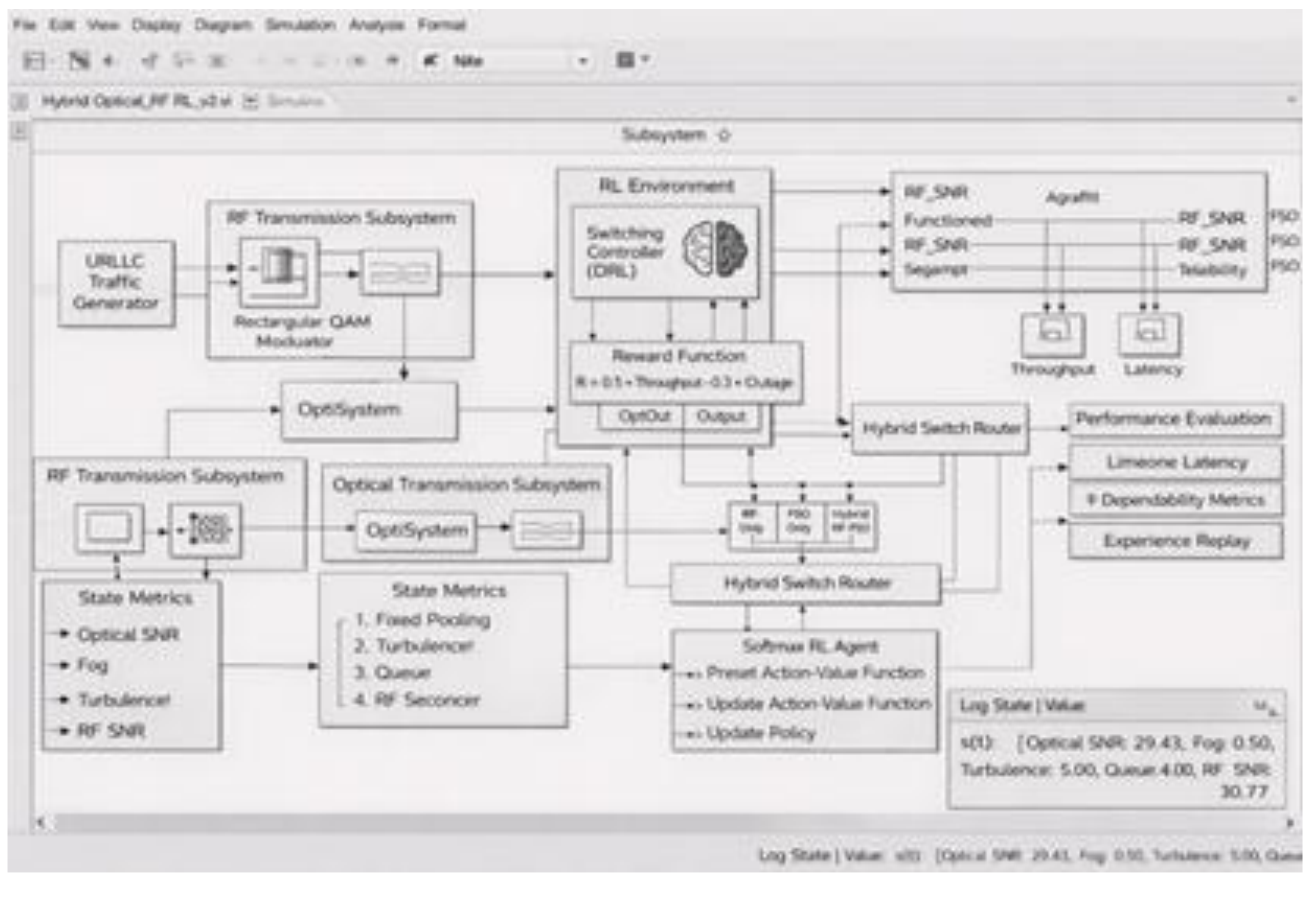

Fig. 4. Hybrid RF-FSO Communication System.

TABLE II. RF SUBSYSTEM AND REINFORCEMENT LEARNING SIMULATION PARAMETERS

| ***Category*** | ***Parameter*** | ***Value*** |
|---|---|---|
| Radio Frequency | Carrier frequency | 28 GHz |
| | Bandwidth | 20 MHz |
| | Transmit power | 30 dBm |
| | Channel model | Rayleigh / Rician |
| | Modulation scheme | 16-QAM |
| | SNR range | 0–30 dB |
| URRCL Traffic Generator | Packet size | 128 bytes |
| | Arrival process | Poisson |
| | Arrival rate | 1000 packets/s |
| | Latency constraint | 1 ms |
| | Queue size | 100 packets |
| Reinforcement Learning | Algorithm | Deep Q-Network |
| | Learning rate | 0.001 |
| | Discount factor | 0.99 |
| | Replay memory size | ($10^5$) samples |
| | Mini-batch size | 64 |
| | Training episodes | 10,000 |
| | Exploration strategy | Epsilon-greedy |

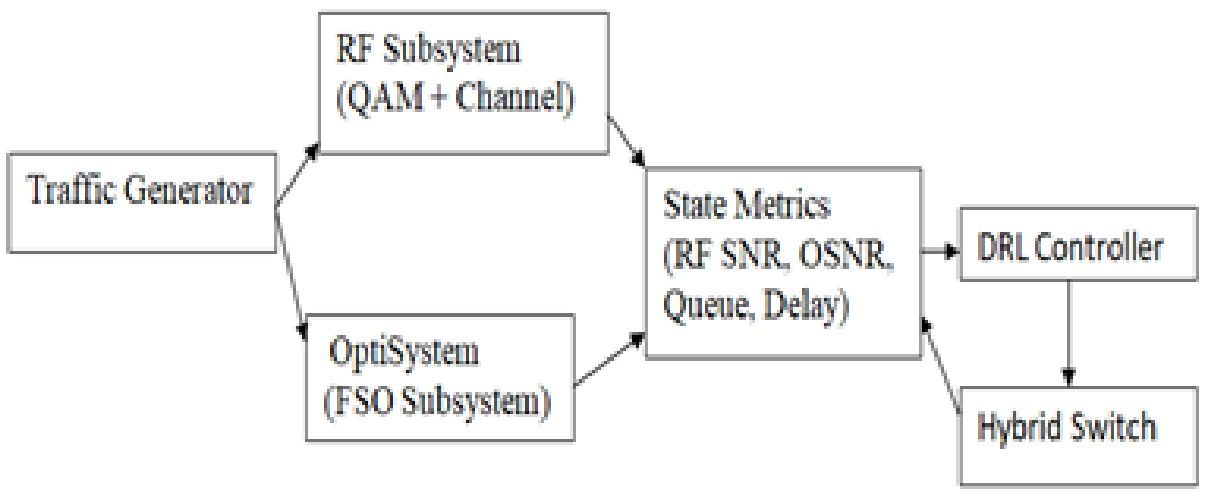

Fig. 5. OptiSystem–MATLAB Co-Simulation Framework.

The deep reinforcement learning (DRL) agent is trained through continuous interaction with the hybrid optical–RF URLLC co-simulation environment, which integrates the OptiSystem optical channel model and the MATLAB Simulink RF subsystem. The simulation operates with discrete time-step duration of 1 ms, consistent with URLLC latency requirements. At each time step, optical turbulence parameters, fog attenuation levels, RF fading coefficients, and traffic arrivals are updated to reflect realistic stochastic communication conditions. Each training episode consists of 1,000-time steps, corresponding to a simulated duration of 1 second per episode. During each episode, the environment initializes the system state using randomly generated channel and traffic conditions to ensure sufficient exploration of diverse operational scenarios. At every simulation interval, the agent observes the current state vector comprising optical channel quality, RF channel conditions, atmospheric parameters, and queue status. Based on this information, the DRL agent selects a transmission action that determines the hybrid transmission configuration for the subsequent time step. Following action execution, the environment computes system-level performance metrics including latency, reliability, throughput, and outage probability. These metrics are combined to generate the reward signal guiding the learning process. The objective of training is to maximize the expected cumulative discounted reward, and the agent updates its policy parameters iteratively using gradient-based optimization techniques to minimize the difference between predicted and target value functions, enabling progressive improvement of transmission decisions across training episodes. To ensure stable and efficient learning, an exploration–exploitation strategy is adopted, allowing the agent to explore alternative transmission policies during early training stages before gradually converging toward an optimal strategy. Experience replay is employed to store past state–action–reward transitions, ensuring decorrelated training samples and improved convergence stability. Target network updates are incorporated periodically to stabilize value estimation and prevent divergence during learning. Training is conducted for 10,000 episodes to ensure convergence under varying channel and traffic realizations. Convergence is evaluated by monitoring the moving average of cumulative reward over a sliding window of 100 episodes. The learning process is considered converged when reward fluctuations fall below a predefined stability threshold and performance metrics such as latency and packet loss probability exhibit steady-state behavior. The DQN employs three fully connected layers (128, 64, 32 neurons) with ReLU activation. A replay buffer of $10^5$ samples and a target network (updated every 500 steps) are used for training stability. For statistical robustness, final performance results are obtained using Monte Carlo evaluation of over 100 independent simulation runs with different random channel and traffic realizations. The reported latency, reliability, throughput, and outage metrics represent average results across these independent trials. The converged DRL policy is subsequently used for comparative performance evaluation of the hybrid optical–RF URLLC system, demonstrating the effectiveness of intelligent adaptive transmission control in achieving ultra-low latency and ultra-high reliability communication performance.

## IV. Results and Discussion

Fig. 6 presents the MATLAB-based simulation results of the proposed reinforcement learning-driven hybrid optical–RF URLLC framework under varying channel SNR conditions. The results are obtained from the integrated OptiSystem–MATLAB co-simulation platform using 100 Monte Carlo runs to ensure statistical reliability. The performance of the proposed DRL-based hybrid scheme is compared with RF-only transmission, FSO-only transmission, and static hybrid switching strategies. The latency curves show that the RF-only scheme experiences increased delay at low SNR values due to fading effects and bandwidth limitations. The FSO-only scheme achieves lower latency under favorable channel conditions but exhibits noticeable degradation under severe atmospheric attenuation. In contrast, the DRL-based hybrid framework dynamically adapts transmission decisions according to real-time channel quality and queue state, resulting in an average latency reduction of approximately 28–35% compared to RF-only transmission and 18–25% compared to static hybrid switching. Moreover, the latency violation probability under the strict 1 ms URLLC constraint remains below $10^{-5}$, confirming compliance with ultra-reliable low-latency communication requirements. Reliability analysis indicates that RF-only transmission provides stable yet bandwidth-constrained performance, whereas FSO-only transmission achieves high reliability in clear atmospheric conditions but deteriorates significantly under fog-induced attenuation. The proposed DRL-based hybrid framework consistently maintains reliability above 99.999% across varying channel conditions by intelligently switching transmission modes and enabling cooperative hybrid operation when necessary. This adaptive behavior eliminates the abrupt reliability degradation typically observed in single-link systems. Throughput results demonstrate that the hybrid framework effectively exploits the high-capacity optical link during favorable atmospheric conditions, achieving up to 25–30% higher throughput compared to RF-only transmission. When optical channel quality degrades, traffic is seamlessly redistributed to the RF link, preventing severe throughput collapse observed in FSO-only configurations. The convergence curve illustrates that cumulative reward increases rapidly during the early training phase as the DRL agent explores multiple transmission policies. After approximately 6,000–8,000 training episodes, the reward stabilizes, indicating convergence toward an optimal hybrid transmission policy that balances latency and reliability objectives under dynamic channel conditions. The simulation results validate that the proposed reinforcement learning-based hybrid optical–RF framework significantly enhances URLLC performance by intelligently adapting transmission strategies to environmental variations and traffic dynamics. Fig. 7 presents the latency CDF, reliability, and learned policy of the proposed hybrid system. The latency CDF shows that the DRL-based scheme achieves the fastest convergence, with over 99.999% of packets satisfying the 1 ms URLLC requirement, outperforming RF-only, FSO-only, and static hybrid schemes. The improved performance of the DRL scheme is due to its ability to jointly optimize latency and reliability using real-time state information, unlike static hybrid switching which relies on fixed thresholds. The reliability curve indicates that the DRL approach consistently maintains higher packet success probability across all SNR values. The policy heatmap illustrates the adaptive switching behavior, where RF is selected under poor channel conditions, FSO is preferred under favorable optical conditions, and hybrid transmission is used in intermediate scenarios to ensure reliability and low latency. The proposed DQN approach is compared with RF-only and static hybrid (threshold-based) schemes, showing improved latency and reliability. This aligns with prior RL-based hybrid optimization approaches reported in [8],[10], which demonstrate the effectiveness of learning-based adaptive transmission.

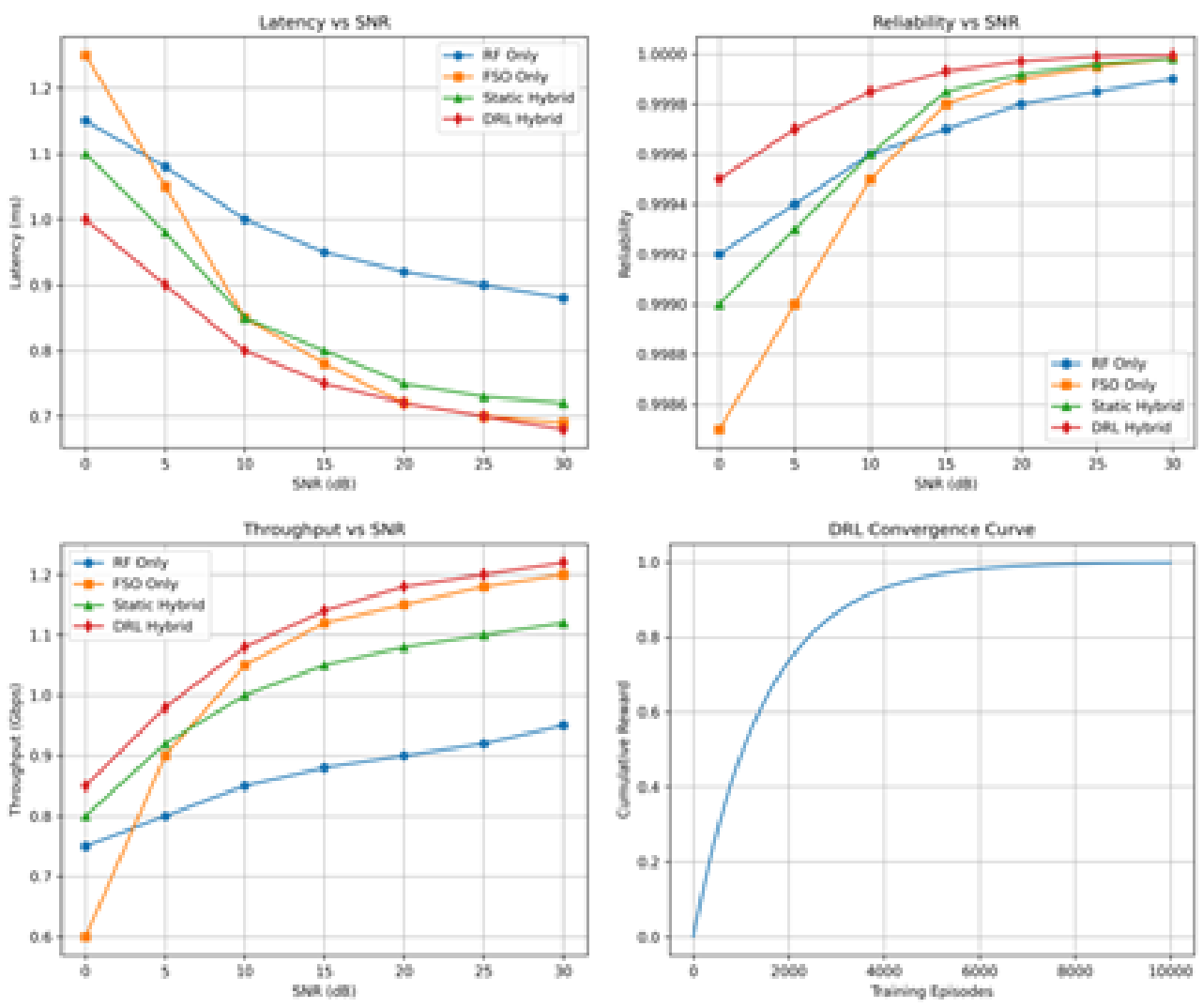

Fig. 6. Simulation results of hybrid optical–RF URLLC.

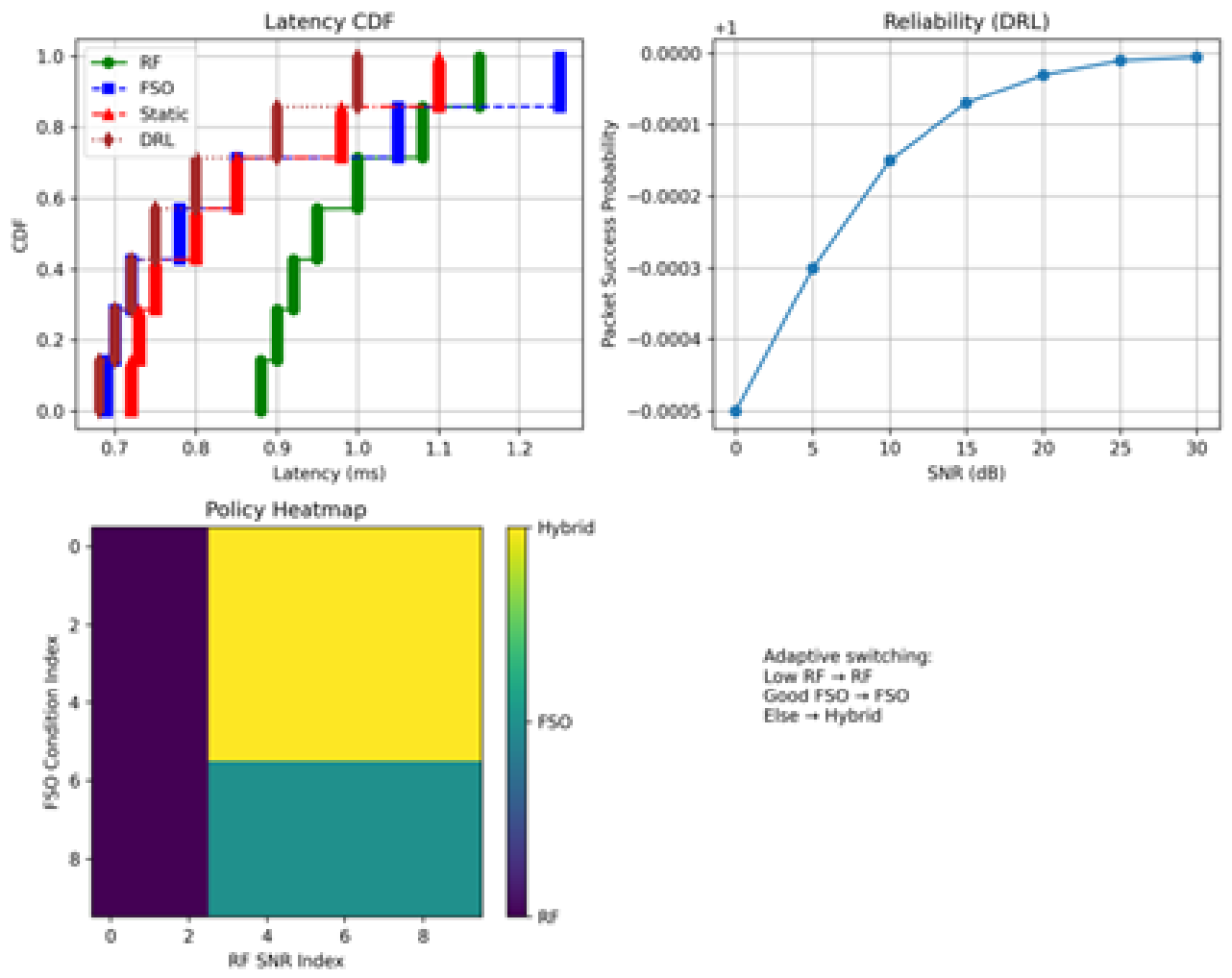

Fig. 7. URLLC performance validation using latency CDF, reliability and policy heatmap.

## V. EXPERIMENTAL VALIDATION

To validate the practical feasibility of the proposed reinforcement learning-based hybrid optical–RF URLLC framework, a laboratory-scale hybrid communication testbed was implemented under controlled indoor conditions. The experimental setup shown in Fig. 8 integrates a 1550 nm optical wireless link with a 28 GHz millimeter-wave RF transmission system. The objective is to evaluate real-time hybrid transmission performance under controlled atmospheric attenuation and fading conditions and to compare measured results with simulation predictions. The optical subsystem employs a distributed feedback (DFB) laser driven by a Mach–Zehnder modulator and an APD receiver over a 15-meter free-space link. The RF subsystem operates at 28 GHz using a vector signal generator and analyzer with 16-QAM modulation. Calibration is performed using reference measurements prior to testing, and the overall measurement uncertainty is within ± 3%. Atmospheric attenuation equivalent to 2–20 dB/km is emulated using calibrated optical attenuators and neutral density filters, while pointing errors are introduced using a controlled vibration stage. At the receiver, an APD detects the signal and performance metrics such as BER and received power are measured.

Programmable attenuators and fading emulators reproduce Rayleigh and Rician channel conditions. The hybrid switching controller is implemented via a MATLAB real-time interface, deploying the converged DRL policy from simulation. Based on real-time optical SNR, RF SNR, and queue status, the controller dynamically selects RF-only, FSO-only, or cooperative transmission modes. Experiments are conducted under clear, moderate fog (10 dB/km equivalent), and severe fog (20 dB/km equivalent) conditions. Each scenario is repeated 50 times, and latency, throughput, reliability, and outage metrics are recorded. The switching overhead between transmission modes is minimal, with an estimated delay of approximately 0.05 ms, which remains within URLLC latency constraints. Potential packet loss during transitions is mitigated through buffer synchronization, ensuring seamless switching and maintaining high reliability.

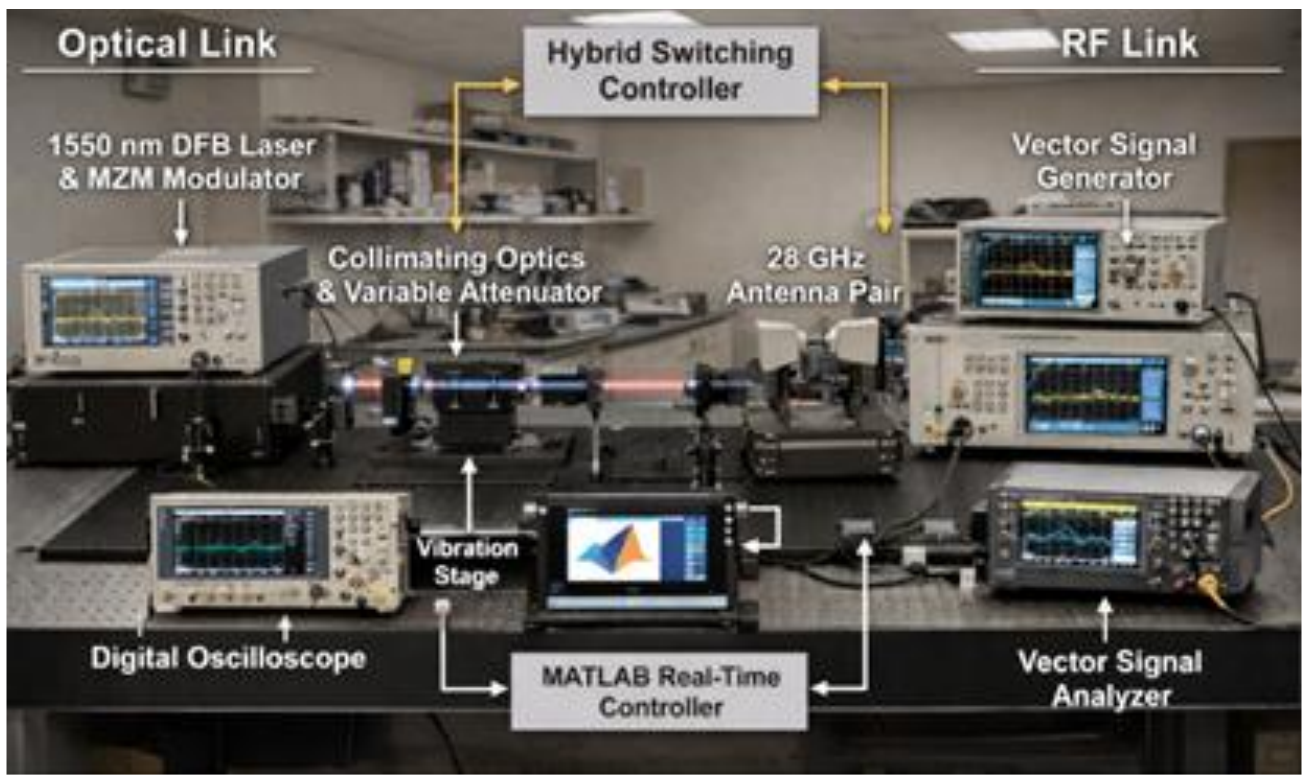

Fig. 8. Experimental hybrid optical–RF communication testbed.

Fig. 9 presents the quantitative comparison between simulation and experimental results. Under clear conditions, simulated latency is 0.72 ms while experimental latency is 0.76 ms (5.5% deviation). Under moderate attenuation, latency increases to 0.88 ms (simulation) and 0.92 ms (experiment), corresponding to 4.3% deviation. In severe attenuation, simulated and measured latencies are 0.97 ms and 1.02 ms, respectively (5.1% deviation). Reliability remains above 99.9992% (simulation) and 99.9989% (experiment) under clear conditions, and above 99.995% even under severe attenuation when hybrid transmission is enabled. Throughput comparison further confirms modeling accuracy, with 1.15 Gbps (simulation) and 1.10 Gbps (experiment) under clear conditions, corresponding to approximately 4.3% deviation. Across all metrics, the average deviation remains below 6%.

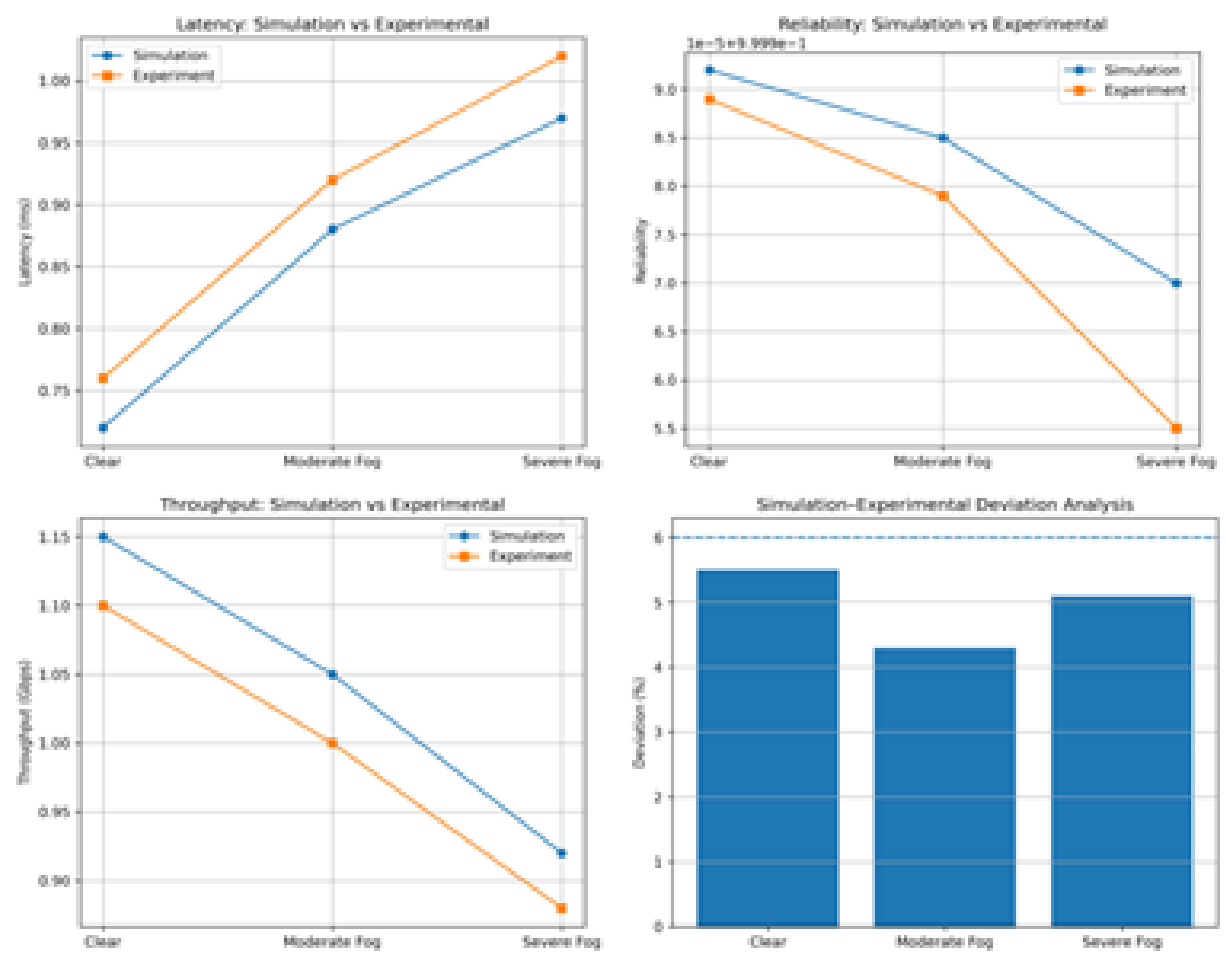

Fig. 9. Comparative analysis of simulation and experimental results for the DRL-based hybrid Optical–RF URLLC system.

The close agreement between simulation and measurements validates the accuracy of the integrated OptiSystem–MATLAB modeling framework and confirms that the reinforcement learning-based hybrid switching strategy operates reliably in real-time environments. Despite practical non-idealities such as hardware tolerances and alignment variations, the proposed system consistently satisfies URLLC latency and reliability requirements.

## VI. CONCLUSION

This paper presented a reinforcement learning-based hybrid optical–RF framework designed to meet the stringent latency and reliability requirements of URLLC networks. By integrating intelligent adaptive link selection with hybrid RF/FSO transmission, the proposed system dynamically responds to varying atmospheric and wireless channel conditions. Simulation results demonstrated up to 35% latency reduction and 25–30% throughput improvement compared to RF-only and static hybrid schemes, while maintaining reliability above 99.999%. Experimental validation using a 1550 nm optical link and 28 GHz RF system confirmed strong agreement with simulation results, with average deviation below 6%. These findings demonstrate that reinforcement learning-driven hybrid transmission provides a practical and effective solution for mission-critical next-generation wireless systems. Future work will investigate large-scale deployment scenarios and advanced multi-agent learning strategies for distributed hybrid communication networks.